\documentclass[12pt]{article}
\usepackage{amsfonts}
\usepackage{psfig}

\newcommand{\beq}{\begin{equation}}
\newcommand{\eeq}{\end{equation}}
\newcommand{\beqn}{\begin{eqnarray}}
\newcommand{\eeqn}{\end{eqnarray}}
\newcommand{\bea}[1]{\beq\begin{array}{#1}}
\newcommand{\eea}{\end{array}\eeq}

\newcommand{\dual}[1]{{}^{*}{#1}}

\newcommand{\tr}{\mathop{\rm Tr}}

\newcommand{\Pexp}{\mbox{P}\!\exp}

\newcommand{\ket}[1]{|\,#1\,\rangle}
\newcommand{\bra}[1]{\langle\,#1\,|}
\newcommand{\braket}[2]{\langle\,#1\,|\,#2\,\rangle}

\newcommand{\diff}{\partial}
\newcommand{\cC}{{\cal C}}

\newcommand{\cS}{{\cal S}}

\newcommand{\NP}[3]{{\it Nucl. Phys. }{\bf #1} (#2) #3}
\newcommand{\NPPS}[3]{{\it Nucl. Phys. Proc. Suppl. }{\bf #1} (#2) #3}
\newcommand{\PL}[3]{{\it Phys. Lett. }{\bf #1} (#2) #3}
\newcommand{\PRL}[3]{{\it Phys. Rev. Lett. }{\bf #1} (#2) #3}
\newcommand{\PRep}[3]{{\it Phys. Rep. }{\bf #1} (#2) #3}
\newcommand{\PR}[3]{{\it Phys. Rev. }{\bf #1} (#2) #3}
\newcommand{\JL}[3]{{\it JETP Lett. }{\bf #1} (#2) #3}
\newcommand{\CMP}[3]{{\it Comm. Math. Phys. }{\bf #1} (#2) #3}

\newcommand{\PTPS}[3]{{\it Prog. Theor. Phys. Suppl. }{\bf #1} (#2) #3}

\newcommand{\JMP}[3]{{\it J. Math. Phys. }{\bf #1} (#2) #3}
\newcommand{\JGP}[3]{{\it J. Geom. and Phys. }{\bf #1} (#2) #3}
\newcommand{\JP}[3]{{\it J. Phys. }{\bf #1} (#2) #3}
\begin{document}
\date{}
\title{The Berry Phase and Monopoles\\
in Non-Abelian Gauge Theories.
\vskip-40mm
\rightline{\small ITEP-TH-14/00}
\vskip 40mm
}
\author{F.V.~Gubarev$^{\rm a,b}$, V.I.~Zakharov$^{\rm b}$ \\
\\
$^{\rm a}$ {\small\it Institute of Theoretical and  Experimental Physics,}\\
{\small\it B.Cheremushkinskaya 25, Moscow, 117259, Russia}\\
$^{\rm b}$ {\small\it Max-Planck Institut f\"ur Physik,}\\
{\small\it F\"ohringer Ring 6, 80805 M\"unchen, Germany}
}

\maketitle
\thispagestyle{empty}
\setcounter{page}{0}
\begin{abstract}\noindent
We consider the quantum mechanical notion of the geometrical (Berry) phase
in SU(2) gauge theory, both in the continuum and on the lattice. It is shown
that in the coherent state basis eigenvalues of the Wilson loop operator
naturally decompose into the geometrical and dynamical phase factors. Moreover,
for each Wilson loop there is a unique choice of U(1) gauge rotations which
do not change the value of the Berry phase. Determining this U(1) locally in
terms of infinitesimal Wilson loops we define monopole-like defects and study
their properties in numerical simulations on the lattice. The construction is
gauge dependent, as is common for all known definitions of monopoles. We argue
that for physical applications the use of the Lorenz gauge is most appropriate.
And, indeed, the constructed monopoles  have the correct continuum limit in
this gauge. Physical consequences are briefly discussed.
\end{abstract}

\newpage
\subsection*{Introduction}
\noindent
In his seminal paper \cite{Berry-original}, Berry found that a quantum state
acquires an extra phase factor when the Hamiltonian of the system depends on
external parameters and the state is adiabatically transported along a closed
path in the parameter space. The additional phase factor depends only on this
path and reflects the geometry of the space of parameters. The mathematical
construction behind the geometrical Berry phase was discovered by Simon \cite{Simon},
who showed that the adiabatic evolution of the quantum system naturally defines
a line bundle structure over the parameter space. The geometrical phase is the
holonomy of the U(1) connexion and cannot be removed by the global redefinition
of the wave function when the bundle is non trivial. Moreover, the adiabatic
approximation can be avoided  and in fact any evolution of a quantum system
naturally gives rise to the geometrical phase factor~\cite{Aharonov-Anandan}
(for a detailed discussion of the non adiabatic Berry phase see \cite{Moore-91}
and references therein). The Berry phase is again the holonomy of the U(1) connexion,
but the base space of the bundle in question is now the space of all physical states.
A unified consideration of the adiabatic and non adiabatic geometrical phases was
presented in \cite{Mostafazadeh-etal,Mostafazadeh}, where it was shown that these
two, a priori different phase factors are actually related by the classification
theorem for vector bundles. Note that the geometrical phase in quantum mechanics
is ultimately related to the construction of the Abelian monopole since this is the
only topologically non trivial object which arises when the structure group is U(1).

In this paper we apply the notion of the Berry phase in the SU(2) pure gauge
theory\footnote{
The SU(2) gauge group is taken for simplicity, the generalization to the SU(N) group is
straightforward.
} and, in particular, in the lattice formulation of the SU(2) gluodynamics. It is known
that the geometrical phase construction is useful in chiral gauge theories (see, e.g.,
\cite{Gauge}). Here we consider the Wilson loop operator in pure gluodynamics from the
quantum mechanical point of view. Indeed, the P--exponent of the gauge potential is
defined as a solution of the first order differential equation, which is very similar to
the Schr\"odinger equation. Therefore the Wilson loop  may be interpreted as the evolution
operator of a quantum mechanical system, where the role of the time-dependent Hamiltonian
is played by the gauge potential. Since the Wilson loop is an operator in the representation
space of the SU(2) group, the space of the physical states is naturally described by the
generalized coherent states \cite{Perelomov}, which parameterize the coset space $SU(2)/U(1)$.
We show that in the coherent state basis eigenvalues of the Wilson loop naturally decompose
into dynamical and geometrical phase factors. For an infinitesimal Wilson loop we obtain the
geometrical phase in terms of the continuum gauge potentials and argue that it reflects 
the non triviality of the Hopf bundle $SU(2)\to SU(2)/U(1)=S^2$.

Our considerations allow to define monopole-like defects in the SU(2) gluodynamics, which
are different from the Abelian monopoles considered so far (see \cite{review} for a review).
Indeed, the usual definition of Abelian monopoles implies a particular partial gauge fixing
which leaves a U(1) subgroup unfixed and the monopoles are defined with respect to this
remaining U(1). While in the standard treatment the U(1) is universally defined for all the
monopoles, in our construction the relevant U(1) subgroup is determined locally for each
infinitesimal Wilson loop or, in the lattice formulation, for each plaquette. 
Therefore, the monopoles constructed are strictly speaking non-Abelian objects.
The existence of the $U(1)$ group inherent for each plaquette was in fact noticed in
Ref.~\cite{ChGuPoZa-00} and is central for the construction developed here.

In particular, the consideration of the geometrical phase gives a new method to calculate
the trace of the Wilson loop, which is periodic function of the phase angle,
${1\over 2}\tr W = \cos\varphi$. We propose a method to calculate the angle $\varphi$
without the restriction $\varphi\in [-\pi ; \pi]$ and introduce the natural decomposition:
\beq
\label{1}
\varphi = \varphi\mathrm{~mod~}2\pi + 2\pi k \,, \qquad\qquad k\in Z \,.
\eeq
The Eq.~(\ref{1}), when applied to the elementary plaquette, gives a lattice analog of the
Dirac strings associated with monopoles we have constructed. Therefore it becomes possible
to consider the non-perturbative dynamics of these defects on the lattice.

An evident problem, which can be understood already in terms of the quantum mechanical analogy,
is that the monopoles constructed in this way are not SU(2) gauge invariant. Following
Ref.~\cite{GuStZa-99} we argue that the most appropriate gauge is the Lorenz gauge, in which
gauge dependent singularities of the gauge fields are maximally suppressed. By numerical
calculations we show that the monopoles we have constructed are physical objects.
In particular, the density of these monopoles is shown to scale correctly towards the
continuum limit. We show that the asymmetry of the monopole currents behaves as an order
parameter of the deconfinement phase transition. Finally, it is shown that these monopoles
are locally correlated with the non-Abelian action density, like the monopoles in the
Maximal Abelian gauge, Ref.~\cite{BaChPo-97}.

The outline of the paper is as follows. In Section~1 we briefly review some  well known
facts about the adiabatic and non adiabatic Berry phases in quantum mechanics. In Section~2
we consider the Wilson loop operator in the SU(2) gluodynamics. In Section~3 the SU(2)
lattice gauge theory is considered. In particular, we present an explicit calculation
of the Berry phase for an arbitrary Wilson loop on the lattice. In Section~4 we discuss
the results of the numerical simulations. Our conclusions are summarized in the last section.

\subsection*{1. Berry Phase in Quantum Mechanics}
\noindent
Following the original paper of Berry~\cite{Berry-original}, consider a hermitian Hamiltonian
$H(t)=H(\lambda_i(t))$ which is time-dependent through a set of parameters $\lambda_i(t)$,
$i=1,...,n$. We are interested in evolution of the quantum mechanical system $\ket{\psi(t)}$
governed by the Schr\"odinger equation:
\beq
\label{Schrodinger-1}
i\diff_t \ket{\psi(t)} ~=~ H(\lambda(t)) ~ \ket{\psi(t)}
\eeq
on the time interval $[0;T]$. Suppose, that for any fixed $t\in [0;T]$ the spectrum of $H$
is non-degenerate:
\beq
H(\lambda) ~ \ket{n(\lambda)} ~=~ E_n(\lambda) ~ \ket{n(\lambda)} \qquad E_n \neq E_m
\eeq
and that the time evolution of the parameters $\lambda(t)$ is slow, such that there are no
induced transitions between energy levels. Then the adiabatic approximation applies and
if the quantum mechanical system at $t=0$ is described by an eigenstate of Hamiltonian
$\ket{\psi(0)} = \ket{n(\lambda(0))}$, the whole evolution reduces to a phase factor:
\beq
\ket{\psi(t)} ~=~ e^{i\varphi(t)}\;\ket{n(\lambda(t))} \,.
\eeq
Moreover, the phase factor $\varphi$ acquired by the state $\ket{\psi(t)}$ as a result of
the adiabatic evolution consists of two parts \cite{Berry-original}:
\beq
\label{phase-factor}
\ket{\psi(t)} ~=~ e^{i\gamma_n(t)} \; \exp\{ -i \int\limits_0^t E_n(\tau) d\tau \}\;
\ket{n(\lambda(t))}\,,
\eeq
where $\delta = - \int E_n $ is the dynamical phase while the additional phase factor
$\gamma_n$ is known as the adiabatic Berry phase. The latter can be found by substituting
(\ref{phase-factor}) into the Schr\"odinger equation (\ref{Schrodinger-1}):
\beq
\diff_t \gamma_n ~=~ i \bra{n(\lambda)} {\diff \over \diff\vec{\lambda}} \ket{n(\lambda)}
\; \diff_t \vec{\lambda}\,.
\eeq

Consider now the parameters $\lambda_i(t)$ which are $T$-periodic, $\lambda_i(0)=\lambda_i(T)$,
and span a closed curve $\cC$ in the parameter space. Then in the adiabatic approximation the
initial and final state vectors differ only by a phase
\beq
\label{cyclic}
\ket{\psi(T)}=e^{i\varphi(T)}\;\ket{\psi(0)}\,.
\eeq
A state vector which obeys Eq.~(\ref{cyclic}) is called the cyclic state. The Berry phase
$\gamma_n(T)$ acquired by the cyclic state $\ket{\psi}$  may be represented in terms of the
"gauge potential":
\beq
\label{berry-connection}
A_i~=~ i\;\bra{n(\lambda)} {\diff \over \diff\lambda_i} \ket{n(\lambda)}\,,
\eeq
\beq
\label{berry-geometrical}
\gamma_n(T) ~=~ \int\limits_{\cC} A ~=~ \int\limits_{\cS_\cC} F\,\,, \quad\qquad F~=~dA\,,
\eeq
where $\cS_\cC$ is an arbitrary surface in the parameter space bounded by the contour $\cC$.
In the form (\ref{berry-connection},\ref{berry-geometrical}) the geometrical nature of the
Berry phase is almost evident. For this reason the Berry phase factor is also called
geometrical phase. Indeed, the rich mathematical structure behind  (\ref{berry-connection},
\ref{berry-geometrical}) was discovered by Simon \cite{Simon} and further elaborated
in a number of papers \cite{Aharonov-Anandan,Mostafazadeh-etal,Mostafazadeh},
\cite{Page-87}--\cite{Anandan-Stodolsky}. In particular, it was shown in
\cite{Aharonov-Anandan} that the adiabaticity requirement is in fact unnecessary. Moreover,
the precise form of the Hamiltonian is not important either. The geometrical phase factor
depends only on the path $\cC$ in the space of the physical states followed by the vector
$\ket{\psi}$ and is naturally associated with any quantum evolution. For self-consistency
we briefly review the essentials of the construction.

Consider unitary evolution of a quantum mechanical system described by a state vector
$\ket{\psi(t)}$, where $\ket{\psi}$ is an element of an $N+1$-dimensional complex vector
space ${\mathbb C}^{N+1}$ with finite $N$, for simplicity. In terms of the complex coordinates
$\{z_0,...,z_N\}$ the state vector is $\ket{\psi(t)}=\{z_0(t),...,z_N(t)\}$. Since the unitary
evolution preserves the norm, the normalization condition $\braket{\psi}{\psi}=1$ defines a
$2N+1$-dimensional sphere $S^{2N+1}\in{\mathbb C}^{N+1}$ on which the evolution takes place. 
The physical states are not given by normalized vectors in ${\mathbb C}^{N+1}$, since
$\ket{\psi}$ and $\ket{\psi\,'}$ define the same physical state if they differ by a phase
$\ket{\psi} = e^{i\alpha}\,\ket{\psi\,'}$. Therefore, the set of physical states is the
$N$-dimensional projective space which is a K\"ahlerian manifold:
\beq
\label{CPN}
{\mathbb CP}^N ~=~ S^{2N+1}/\,U(1)\,.
\eeq

The quantum mechanical evolution is given by the Schr\"odinger equation (\ref{Schrodinger-1}),
where the Hamiltonian depends implicitly on time through the set of the parameters $\lambda_i(t)$, 
$\lambda_i(0) = \lambda_i(T)$. We are interested in a cyclic vector $\ket{\psi(t)}$ which returns 
to the same physical state $\ket{\psi(T)}= e^{i\varphi(T)}\;\ket{\psi(0)}$ after evolving along
a closed path $\cC \in {\mathbb CP}^N$. The phase $\varphi(T)$ is the total phase acquired by
the cyclic vector. Moreover, we introduce a single-valued vector $\ket{\tilde{\psi}(t)}$
which differs from $\ket{\psi(t)}$ by a phase and satisfies the condition:
\beq
\label{single-valued}
\ket{\tilde{\psi}(T)} ~=~ \ket{\tilde{\psi}(0)}\,.
\eeq
Note that Eq.~(\ref{single-valued}) does not define the $\ket{\tilde{\psi}}$ uniquely:
for a given $\ket{\tilde{\psi}}$,  vector $e^{i\theta}\;\ket{\tilde{\psi}}$ is also
single-valued provided that $\theta(T)-\theta(0)=2\pi n$.

In terms of the single-valued vector $\ket{\tilde{\psi}}$ the cyclic state $\ket{\psi}$
is represented as:
\beq
\ket{\psi(t)} ~=~ e^{i\varphi(t)} \; \ket{\tilde{\psi}(t)}
\eeq
and the Schr\"odinger equation gives:
\beq
\label{nonadiabatic-total}
\varphi(T)~=~ \delta ~+~ \gamma ~=~ - \int\limits_0^T \bra{\tilde{\psi}} H \ket{\tilde{\psi}}
~+~ i \int\limits_\cC \bra{\tilde{\psi}} d \ket{\tilde{\psi}}\,.
\eeq
The first term in Eq.~(\ref{nonadiabatic-total}) is naturally identified with the non adiabatic
dynamical phase and depends explicitly  on the detailed structure of the Hamiltonian. One can
show \cite{Aharonov-Anandan} that in the adiabatic limit it reduces to the expression
$\delta=-\int E_n$ mentioned above. Note that the dynamical phase may also be calculated as 
$\delta = - \int\limits \bra{\psi} H \ket{\psi}$ since the phase difference
between $\ket{\psi}$ and $\ket{\tilde{\psi}}$ drops out in the matrix element.
The second term is a non adiabatic generalization of the geometrical Berry phase
(\ref{berry-geometrical}) and it depends only on the closed path $\cC\in{\mathbb CP}^N$
spanned by $\ket{\tilde{\psi}}$ during its cyclic evolution. Indeed, the geometrical phase
$\gamma\,'$ calculated by means of  any other single-valued vector
$\ket{\tilde{\psi}\,'}= e^{i\theta}\;\ket{\tilde{\psi}}$
(see the note after Eq.~(\ref{single-valued})) differs from $\gamma$, Eq.~(\ref{nonadiabatic-total}):
\beq
\gamma\,' ~=~ i \int\limits_\cC \bra{\tilde{\psi}} \;e^{-i\theta }\;d\;e^{i\theta }\;\ket{\tilde{\psi}}
~=~ \gamma - \int\limits_\cC d\theta ~=~ \gamma -2\pi n
\eeq
and therefore the difference $\gamma\,'-\gamma$ is  inessential.

In order to explicitly evaluate the geometrical phase we note that single-valued vectors
$\ket{\tilde{\psi}}$ are parameterized by the homogeneous coordinates $w_i$ on ${\mathbb CP}^N$,
$\ket{\tilde{\psi}}= \{w_1,...,w_N\}$, and the scalar product $\braket{\tilde{\psi}\,'}{\tilde{\psi}}$
is given by:
\bea{c}
\ket{\tilde{\psi}}= \{w_1,...,w_N\}\,,  \qquad  \ket{\tilde{\psi}\,'}= \{w_1',...,w_N'\}\,,  \\
\\
\braket{\tilde{\psi}\,'}{\tilde{\psi}} ~=~ 
\exp\{ K(\bar{w}',w) - \frac{1}{2} K(\bar{w}',w') - \frac{1}{2} K(\bar{w},w) \}\,,
\eea
where $\bar{w}$ denotes complex conjugation and $K(\bar{w}',w)$ is the K\"ahler potential on
${\mathbb CP}^N$. One obtains \cite{Page-87,Onofri}:
\beq
\label{gamma}
\gamma ~=~ -\int\limits_\cC \mathrm{Im}\left[ {\diff K(\bar{w},w) \over \diff w} dw \right] ~=~
\int\limits_\cC \frac{i}{2}\; {\bar{w}_i dw_i - w_i d\bar{w}_i \over 1 + \bar{w}_i w_i }\,.
\eeq

The appearance of a non trivial geometrical phase during a cyclic evolution of the quantum
mechanical system is guaranteed then by the following topological arguments 
\cite{Mostafazadeh-etal,Mostafazadeh,Kiritsis-87}.
Let variables $\lambda_i$ parameterize a manifold $\Lambda$.
We have shown that the Hamiltonian $H(\lambda)$ defines the mapping $f : \Lambda \to {\mathbb CP}^N$
which maps any curve in $\Lambda$ to a curve in ${\mathbb CP}^N$.
Since $\pi_n({\mathbb CP}^N)=0$, $n \neq 2$ and $\pi_2({\mathbb CP}^N)=\mathrm{Z}$
the mapping $f$ might not be homotopic to zero provided that $\pi_2(\Lambda)\neq 0$.
On the other hand, the line bundle (\ref{CPN}) is non trivial and therefore the mapping
$S^{2N+1}\to {\mathbb CP}^N \stackrel{f^*}{\to} \Lambda$ defines a non trivial U(1) bundle
over $\Lambda$. In this case the global definition of the phase of the state vector
is impossible over parameter space and this fact guarantees the appearance of the Berry phase.
Note that the geometrical phase factor when written in the form:
\beq
\label{integral}
\gamma ~=~ i \int\limits_\cC \bra{\tilde{\psi}} d \ket{\tilde{\psi}} ~=~ \int\limits_\cC A ~=~
\int\limits_{\cS_\cC} F\,,
\eeq
where $\cS_\cC$ is an arbitrary surface which bounds the contour $\cC\in \Lambda$, defines the
winding number density $\frac{1}{2\pi} F \sim \diff_{\lambda_i}\bra{\tilde{\psi}}
\;\, \diff_{\lambda_j}\ket{\tilde{\psi}} \; d\,^2\lambda^{ij}$ of the mapping
$\Lambda\ni S^2 \to {\mathbb CP}^N$, since
$(\mathrm{d}\bra{\tilde{\psi}}) \,\wedge\, (\mathrm{d}\ket{\tilde{\psi}})$ is the first Chern
class of the bundle (\ref{CPN}). In particular, the integral of $ {1\over 2\pi}F$ over
$S^2\in \Lambda$ gives the degree of the mapping $\Lambda\ni S^2 \to {\mathbb CP}^N$.
This implies in turn that the integral (\ref{integral}) calculated
for two different surfaces $\cS_\cC$ and $\cS_\cC'$ may differ only by $2\pi n$ and therefore
the phase factor $e^{i\gamma}$ is surface independent. The non triviality of the mapping
$\Lambda\ni S^2 \to {\mathbb CP}^N$, together with a line bundle structure (\ref{CPN}),
is essentially the only reason why the monopole potential repeatedly appears in studies of
the Berry phase.

Note that both dynamical and geometrical phase factors calculated in unitary transformed
state basis are generally different. Indeed, going to unitary transformed states
$\ket{\tilde{\psi}'} = U \ket{\tilde{\psi}}$ is equivalent to going to the Hamiltonian
$H'(\lambda)= U^+ H(\lambda) U -i U^+ \diff_t U$. The corresponding total phase is then:
\bea{c}
\label{unitary}
\varphi\,' ~=~  - \int\limits_0^T \bra{\tilde{\psi}} U^+ H U \ket{\tilde{\psi}} 
~+~ i \int\limits_\cC \left[ \bra{\tilde{\psi}}U^+ d U \ket{\tilde{\psi}} ~+~
\bra{\tilde{\psi}} d \ket{\tilde{\psi}} \right] ~=
\\
=~ - \int\limits_0^T \bra{\tilde{\psi}'} H \ket{\tilde{\psi}'} ~+~
i \int\limits_{\cC\,'} \bra{\tilde{\psi}'} d \ket{\tilde{\psi}'}\,.
\eea
It is clear that both dynamical and geometrical phases calculated with $H'$ differ in
general from the phases evaluated with $H$. Nevertheless, there exists a class of unitary
transformations $U(t)$ for which the Berry phases are the same. Namely, if for a unitary
$U(t)$ we have $\cC' = \cC$, then  $\gamma\,'=\gamma$ since the geometrical phase depends
only on the path $\cC\in{\mathbb CP}^N$. It is easy to see that only the U(1)
transformations, $U = e^{i\alpha}$ possess this property.

\subsection*{2. Geometrical Phase and Wilson Loop}
\noindent
There is a class of Hamiltonians which are of particular interest and are constructed as follows.
The space of state vectors carries a unitary irreducible re\-pre\-sen\-ta\-tion of a com\-pact
semi\-simple Lie group $G$. The Hamiltonian depends on a set of parameters $\lambda_i(t)$ and
for every $t$ is an element of the Lie algebra of $G$. Then the evolution operator for the
Schr\"odinger equation (\ref{Schrodinger-1}),
\beq
\label{evolution-V}
V(t)~=~ \mathrm{T} \exp\{ -i \int_0^t H\}
\eeq
is given by a path $[0;T]\to G$ in the group space and for any $t$ belongs to the representation
of $G$ (in the Eq.~(\ref{evolution-V})~ $\mathrm{T}$ denotes the time-ordering).
Therefore, if at $t=0$ we start with an arbitrary state $\ket{\psi(0)}$, then the state vector
at the time $t$, $\ket{\psi(t)}=V(t)\ket{\psi(0)}$ is a generalized coherent state \cite{Perelomov}.
It is important that the phase of the state $\ket{\psi(t)}$ is naturally fixed with respect
to a specific choice of the coherent state basis $\ket{z}$. Indeed, an arbitrary state vector may
be represented as $\ket{\psi(t)} = e^{i\varphi(t)} \ket{z(t)} = e^{i\varphi(t)} U(z(t)) \ket{0}$,
where the complex variables $\{z_1,...,z_n\}$  parameterize the coset space $G/H$, $H$ is the
Cartan subgroup of $G$, $U(z)\in G/H$ and $\ket{0}$ denotes an arbitrary fixed vector in the
representation space, which is usually taken to be highest or lowest weight vector.
Thus the space of all physical states is nothing else but the coset space $G/H$.
The total phase factor acquired by the cyclic state during the evolution (\ref{evolution-V})
may be calculated by means of Eq.~~(\ref{nonadiabatic-total}) with $\ket{\tilde{\psi}}=\ket{z}$ 
\cite{Onofri,Maslanka}.

The evolution operator (\ref{evolution-V}) is of particular importance in gauge theories.
Indeed, consider a Wilson loop $W(T)$ on the contour parameterized by coordinates
$x_\mu(t)$, $t\in [0;T]$:
\beq
\label{Wilson-loop}
W(T) ~=~ \Pexp\{ i\int_0^T A(t) \; dt \}\,, \qquad
A(t)~=~ A^a_\mu(x(t)) \;\dot{x}_\mu(t)\; T^a \,,
\eeq
where $T^a$ are the generators of the gauge group $G$ in the representation considered and the
dot denotes differentiation  with respect to $t$.  The P-exponent is  defined as a solution of
the first order differential equations:
\beq
(\; i\diff_t ~+~ A\;) \ket{\psi} ~=~0\,, \qquad
\ket{\psi(t)}~=~ W(t)\;\ket{\psi(0)}
\eeq
and therefore the Wilson loop is just the evolution operator (\ref{evolution-V}) with  a
time-dependent Hamiltonian $H=-A(t)$. Therefore, the discussion above suggests that phases of
eigenvectors of the Wilson loop may be naturally decomposed  into dynamical and geometrical
parts. Note that such a decomposition cannot be gauge invariant. Indeed, we have shown in the
previous section that the dynamical and geometrical phases calculated for unitary equivalent
states vectors are generally different. On the other hand, the transformation 
$H' = U^+ H U - i U^+ \diff_t U$ is exactly the same as the gauge transformation.

In the simplest case of $G=SU(2)$ and the Wilson loop in the fundamental representation, the
procedure is as follows. The space $G/H$ is a two-dimensional sphere which is parameterized
by a complex coordinate $z$. The coherent states are build over the highest weight vector
$\ket{0}$, $\sigma^3\ket{0}=\ket{0}$, $\sigma^a$ are the Pauli matrices:
\beq
\label{coherent-z}
\ket{z}~=~{1\over \sqrt{1+|z|^2}}\; e^{\;z\,(\sigma^1-i\sigma^2)/2} \; \ket{0}\,.
\eeq
The action of the group element $g\in SU(2)$ in the coherent state basis\footnote{
Note that our notations are slightly different (although equivalent) to that of
Ref.~\cite{Perelomov}.
} is \cite{Perelomov}:
\beq
\label{action-on-coherent-states-1}
g~=~
\left[\begin{array}{cc}
\alpha & \beta \\
-\bar{\beta} & \bar{\alpha}
\end{array}\right]\,,
\qquad\qquad
g\ket{\zeta}~=~e^{i\phi(g,\zeta)}\;\ket{\zeta_g}\,,
\eeq
\beq
\label{action-on-coherent-states-2}
\phi(g,\zeta) ~=~ \mathrm{arg}[\beta\zeta+\alpha]\,,
\qquad\qquad
\zeta_g ~=~ { \bar{\alpha}\zeta - \bar{\beta} \over \beta\zeta + \alpha}\,.
\eeq
The Wilson loop operator, Eq.~(\ref{Wilson-loop}), has two eigenstates $\ket{z_{\pm}}$:
\beq
W(t)~=~
\left[\begin{array}{cc}
\alpha(t) & \beta(t) \\
-\bar{\beta}(t) & \bar{\alpha}(t)
\end{array}\right]\,,
\qquad\qquad
W(T)\ket{z_{\pm}}~=~e^{\pm i\varphi(T)}\;\ket{z_{\pm}}\,.
\eeq
It is sufficient to consider a particular eigenvector, e.g. $\ket{z_+}$, which tends to
$\ket{0}$ when $\beta\to 0$ (the $z_-$ approaches infinity in the limit $\beta\to 0$).
The cyclic state $\ket{\zeta(t)}$ and the single valued vector $\ket{z(t)}$ are
constructed as follows:
\beq
\ket{\zeta(t)} ~=~ W(t)\;\ket{z_+} ~=~ e^{i\varphi(t)}\,\ket{z(t)}\,,
\qquad\qquad
z(t) ~=~ { \bar{\alpha}(t)\,z_+ - \bar{\beta}(t) \over \beta(t)\,z_+ + \alpha(t)}\,.
\eeq
Therefore, the total phase factor $\varphi(T)$ acquired by the cyclic state $\ket{\zeta}$ is
\beq
\label{phase-1}
\varphi(T)~=~ \mathrm{arg}[\beta(T)z_+ +\alpha(T)] ~=~ \delta ~+~ \gamma\,,
\eeq
\beq
\label{phase-2}
\delta = \int\limits_0^T \bra{z} A \ket{z} = \int\limits_0^T \bra{\zeta} A \ket{\zeta} \,,
\qquad
\gamma = i \int\limits_\cC \bra{z} d \ket{z} =
-\int\limits_\cC \mathrm{Im}\left[ {\diff K(\bar{z}, z ) \over \diff z } d z\right]\,,
\eeq
where $K(\zeta,z)~=~ \ln[1+\bar{\zeta}z]$ is the K\"ahler potential on $G/H$. Similar to the
quantum mechanical example, the dynamical phase may be calculated with either the single-valued
vector or with the cyclic state. The trace of the Wilson loop operator is evidently
$\frac{1}{2}\tr W(T) = \cos\varphi(T)$.

\subsection*{3. Lattice Implementation.}
\noindent
Eqs.~(\ref{coherent-z}--\ref{phase-2}) are directly applicable in the lattice gauge
theories (LGT). Consider therefore the Wilson loop operator as it appears on the lattice:
\beq
\label{lattice-wilson-loop}
W ~=~ 
\left[\begin{array}{cc}
\alpha_w & \beta_w \\
-\bar{\beta}_w & \bar{\alpha}_w
\end{array}\right]
~=~ g_N \; g_{N-1}\; ... \; g_1 \,,
	\qquad\qquad
g_i~=~
\left[\begin{array}{cc}
\alpha_i & \beta_i \\
-\bar{\beta}_i & \bar{\alpha}_i
\end{array}\right]\,.
\eeq
Its eigenstate is determined by
\beq
\label{eigenstate}
W \ket{z_+} ~=~ e^{i\varphi}\;\ket{z_+}\,,
\qquad
z_+~=~ {-i\over\beta_w} \left[\; \mathrm{Im}\alpha_w -
\mathrm{sign}(\mathrm{Im}\alpha_w) \sqrt{(\mathrm{Im}\alpha_w)^2 + |\beta_w|^2 } \;\right]\,,
\eeq
where the eigenvector which obeys the condition $z_+ \to 0$ when $\beta_w \to 0$  has been selected.
The calculation of the total phase $\varphi$ is now straightforward. Indeed, according
to Eqs.~(\ref{action-on-coherent-states-1},\ref{action-on-coherent-states-2})
the cyclic $\ket{\zeta_k}$ and single-valued $\ket{z_k}$ vectors are given by:
\bea{rcl}
\label{z-k}
g_k  ~\ket{\zeta_{k-1}} & ~=~ & \ket{\zeta_k} \\
\rule{0mm}{6mm}
g_k ~ \ket{z_{k-1}} & ~=~ & \ket{z_k}~e^{i\varphi_k} \\
\rule{0mm}{6mm}
k=1, ... ,N\,, & &  z_0~=~\zeta_0~=~z_N ~=~ \zeta_N e^{-i\varphi}~=~ z_+ \,.
\eea
Therefore, the total phase factor $\varphi$ of the Wilson loop is calculated as:
\beq
\label{total-phase-lattice}
\varphi ~=~ \sum\limits_{k=1}^N \varphi_k\,.
\eeq
By construction, the total phase $\varphi$ satisfies $\frac{1}{2} \tr W = \cos\varphi$. Moreover,
Eq.~(\ref{total-phase-lattice}) reduces to the well known Abelian expression when all $g_i$
are diagonal.

The decomposition of the total phase (\ref{total-phase-lattice}) into the dynamical and geometrical
parts is more subtle. The point is that such a decomposition is not possible without a particular
parameterization of the Wilson loop $W$ and of the matrices $g_k$ in terms of a  continuous time
variable. We choose a natural parameterization:
\beq
\label{Wilson-loop-lattice-time-dependent}
W(t)~=~\left\{
\begin{array}{lc}
g_1(t)\,, &  0\leq t < 1 \\
... & \\
g_k(t) g_{k-1} ... g_1\,, &  k-1\leq t < k \\ 
... & \\
g_N(t) ... g_1\,, &  N-1\leq t < N \\ 
\end{array}
\right. \,,
\eeq
where the time-dependent matrices $g_k(t)$ are defined as follows: if $g_k=\exp\{iA_k\}$ then
$g_k(t)=\exp\{i A_k (t-k+1)\}$. With the parameterization (\ref{Wilson-loop-lattice-time-dependent}),
the Wilson loop $W(t)$ describes a continuous path in the SU(2) group manifold and the cyclic state
$\ket{\zeta(t)}=W(t) \ket{z_+}$ satisfies the Schr\"odinger equation 
$i\diff_t\ket{\zeta} = -A_k \ket{\zeta}$, $k-1\leq t < k$, $k=1, ... , N$.

Since $g_k(t) A_k = A_k g_k(t)$, we obtain from Eq.~(\ref{phase-2}) the dynamical phase:
\beq
\label{lattice-dynamical-phase}
\delta ~=~ \sum\limits_{k=1}^N \bra{z_{k-1}} A_k \ket{z_{k-1}} ~=~ 
\sum\limits_{k=1}^N \bra{\zeta_{k-1}} A_k \ket{\zeta_{k-1}}\,,
\eeq
where $\ket{z_k}$ are defined by Eq.~(\ref{z-k}) and the matrix elements $\bra{z} A \ket{z}$,
$A=A^a\sigma^a/2$ are given by
\beq
\label{matrix-element}
\bra{z} A \ket{z} ~=~ {1\over 1+|z|^2} \left[
{1\over 2} A^3 (1-|z|^2) ~+~ \mathrm{Re}(zA^-)
\right] \,,
\qquad\qquad
A^- ~=~ A^1-iA^2 \,.
\eeq
Moreover, the Berry phase associated with the Wilson loop (\ref{lattice-wilson-loop}) is
$\gamma = \varphi - \delta$.

Consider the Wilson loop  which is the ordered product of the link matrices $g_{x,\mu}$ taken
along the boundary of the elementary plaquette situated at the point $x$ and directed in the
$\{\mu,\nu\}$ plane:
\beq
W ~=~  g^+_{x,\nu} g^+_{x+\nu,\mu} g_{x+\mu,\nu} g_{x,\mu}\,.
\eeq
Let the area of the plaquette be $\sigma^{\mu\nu}$, then the dynamical phase
$\delta_{x,\mu\nu}\,\sigma^{\mu\nu}$ is:
\beqn
\label{unknown}
\delta_{x,\mu\nu}\,\sigma^{\mu\nu} ~=~ \bra{z_+}~
A_{x,\mu} ~+~  g^+_{x,\mu}  A_{x+\mu,\nu} g_{x,\mu} ~-~
g^+_{x,\mu} g^+_{x+\mu,\nu}  A_{x+\nu,\mu}  g_{x+\mu,\nu} g_{x,\mu} ~- \nonumber \\
\rule{0mm}{5mm}
-~ g^+_{x,\mu} g^+_{x+\mu,\nu} g_{x+\nu,\mu} A_{x,\nu}  g^+_{x+\nu,\mu} g_{x+\mu,\nu} g_{x,\mu}~
\ket{z_+} \,\sigma^{\mu\nu} \,.
\eeqn
Note that the state $\ket{z_+}$ depends on the plaquette considered.
For an infinitesimal Wilson loop  Eq.~(\ref{unknown})
can be expanded in powers of $\sigma^{\mu\nu}$. In the leading order the dynamical phase is given by:
\beq
\label{unknown-2}
\delta_{x,\mu\nu}\,\sigma^{\mu\nu} ~=~ \bra{z_+} F_{\mu\nu} - i [A_\mu, A_\nu]   \ket{z_+} \,\sigma^{\mu\nu}
~=~ \bra{z_+} D_{[\mu} A_{\nu]} \ket{z_+} \,\sigma^{\mu\nu}
\,,
\eeq
where $F_{\mu\nu}$ and $D_\mu$ are the continuum field strength tensor and covariant derivative, respectively:
\beq
F_{\mu\nu} = \diff_{[\mu} A_{\nu]}-i[A_\mu, A_\nu] = 
\frac{1}{2}\sigma^a \left( \diff_{[\mu} A^a_{\nu]} + \varepsilon^{abc} A^b_\mu A^c_\nu \right)\,,
\qquad
D_\mu = \diff_\mu - i [A_\mu,\;] \,.
\eeq
It can be checked directly that the total phase factor is
$\varphi_{x,\mu\nu}\,\sigma^{\mu\nu} = \bra{z_+} F_{\mu\nu} \ket{z_+} \,\sigma^{\mu\nu}$.
Therefore, the Berry phase for the infinitesimal Wilson loop is given by:
\beq
\label{Berry-phase-lattice}
\gamma_{x,\mu\nu}\,\sigma^{\mu\nu} ~=~ i \bra{z_+} [A_\mu, A_\nu] \ket{z_+} \,\sigma^{\mu\nu}\,.
\eeq
As expected from the quantum mechanical example above, the value of the Berry phase measures to what
extent the gauge fields are non-Abelian. Indeed, for pure Abelian potentials $A^a_\mu=\delta^{a,3}A_\mu$
the Berry phase vanishes and the total phase of the infinitesimal Wilson loop is given by the
dynamical phase alone:
\beq
\varphi_{x,\mu\nu}~=~ \delta_{x,\mu\nu} ~=~ \frac{1}{2} \diff_{[\mu} A_{\nu]}\,.
\eeq

As is mentioned above, the decomposition of the total phase of the Wilson loop into the dynamical
and geometrical parts, Eqs.~(\ref{phase-2}, \ref{lattice-dynamical-phase}, \ref{unknown-2},
\ref{Berry-phase-lattice}), is gauge dependent. Nevertheless, it is possible to find a subgroup
of the gauge group which does not change the Berry phase. Indeed, the defining property of these
gauge transformations is the same as in the quantum mechanics, namely, they should not change the
path spanned by the state vector $\ket{z(t)}$  in  $G/H ~=~ SU(2)/U(1)$. It is clear that the
subgroup in question is a U(1) subgroup of SU(2). Since $1/2\tr W = \cos\varphi$, the quantity
$\varphi\mathrm{~mod~}2\pi$ is gauge invariant and the effect of the U(1) gauge transformations
which do not change the Berry phase, is to shift the dynamical or, equivalently, total phase 
by $2\pi k$, $k\in Z$. It is amusing to note that with $k \neq 0$ these are indeed "large" U(1) gauge
transformations familiar from the example of the U(1) compact electrodynamics.

Moreover, when the gauge is fixed the considerations of the Section 1 are applicable. Namely, any closed
path $\cC$ in the physical space $M$ is naturally mapped into a closed path in $G/H$, 
the relevant mapping being determined by the Wilson loop operator, calculated on $\cC$.
The degree of the mapping $f : \cS \to G/H$, where $\cS$ is a closed two-dimensional surface in $M$,
is given by:
\beq
\label{degree}
\mathrm{deg}[f] ~=~ {1\over 2\pi} \int\limits_\cS \; \gamma 
~=~ {1\over 2\pi} \int\limits_{G/H} \; F \,\,,
\eeq
where $F$ is the first Chern class of the non trivial Hopf bundle $G\to G/H$. Note that when
$\mathrm{deg}[f]\ne 0$ the Berry phase $\gamma$ contains a string-like singularity (Dirac string)
somewhere on the surface $\cS$. As noted above, the position of the singularity may be
arbitrary shifted by U(1) gauge transformations.

Let the surface $\cS$ be parameterized by coordinates $\tilde{x}(\sigma)$,
$\sigma=\{\sigma^\alpha\}=\{\sigma^1,\sigma^2\}$. Using Eqs.~(\ref{eigenstate}, \ref{matrix-element})
it is straightforward to show that:
\beq
\int\limits_{\cS} \gamma ~=~
-{1\over 2} \int\limits_\cS \,\varepsilon^{abc}\,n^a(\sigma)\,A^b_\mu(\tilde{x})\,A^c_\nu(\tilde{x})\,
\mathrm{d}^2 \sigma^{\mu\nu}
\qquad
\mathrm{d}^2 \sigma_{\mu\nu} = \varepsilon^{\alpha\beta}\diff_\alpha\tilde{x}_\mu \diff_\beta\tilde{x}_\nu\,,
\eeq
where the world-sheet vector field $n^a(\sigma)$, $\vec{n}^2=1$ was introduced in Ref.~\cite{ChGuPoZa-00}:
\beq
\label{n-a}
n^a(\sigma) ~=~ (\mathrm{d}^2 \sigma \cdot F^a ) \left[{(\mathrm{d}^2 \sigma \cdot F^b )^2}\right]^{-1/2}\,,
\qquad
(\mathrm{d}^2 \sigma \cdot F^a ) = \mathrm{d}^2\sigma^{\mu\nu} F^a_{\mu\nu}\,.
\eeq
Note, however, that $n^a(\sigma)$, Eq.~(\ref{n-a}), is build on the $F^a_{\mu\nu}$, not
$\dual{F}^a_{\mu\nu}= \frac{1}{2} \varepsilon_{\mu\nu\lambda\rho} F^a_{\lambda\rho}$
as in Ref.~\cite{ChGuPoZa-00}.

\subsection*{4. Numerical Results.}
\noindent
In this section we show that the formalism presented  above may be used  in lattice simulations
and present the numerical results which concern the total phase, Eq.~(\ref{total-phase-lattice}).
The total phase of the Wilson loop is of particular importance since it naturally includes
the contribution of both Abelian-like and non-Abelian fields.
As for the lattice implementation of the dynamical and Berry phases, we have not considered it yet
although it seems straightforward.

The trace of the Wilson loop, calculated over the boundary of an elementary plaquette $p$, 
is gauge invariant and is proportional to $\cos\varphi_p$.
Therefore, the total phase $\varphi$ may be represented in terms of the gauge invariant
quantity $\varphi_p \mathrm{~mod~}2\pi$ and an integer number $k_{\dual{p}}$:
\beq
\label{lattice-Dirac-string}
\varphi_p ~=~ \varphi_p \mathrm{~mod~}2\pi ~+~ 2\pi k_{\dual{p}} \,,
\eeq
where $\dual{p}$ is the plaquette dual to $p$.
By analogy with the corresponding equality in the compact U(1) LGT the integer $k_{\dual{p}}$
counts the number of Dirac strings piercing the plaquette $p$. A set of numbers $k_{\dual{p}}$
is an integer valued 2--form $k \in \dual{C}_2(Z)$, which belongs to
oriented plaquettes on the dual lattice. As usual, we define the monopoles as end points 
of the Dirac strings (\ref{lattice-Dirac-string}):
\beq
\label{j}
j_{x,\mu}~=~ (\delta k )_{x,\mu}\,,
\qquad \qquad
j\in \dual{C}_1(Z) \,.
\eeq
The monopole current $j_{x,\mu}$ is an integer valued 1--form on the dual lattice.
While all the definitions above are unique in terms of the potentials (or link matrices), the potentials
themselves are gauge dependent. As a result, the values of $k_{*p}$ are gauge dependent as well and at first
sight the definition ({\ref{j}) is devoid of any physical meaning. Clearly, the issue of the  gauge
dependence is crucial for physical applications of the procedure developed above.

Therefore we introduce at this point a physically motivated gauge following the logic similar to that of
the paper \cite{GuStZa-99}. The observation is that in the continuum limit both the Dirac strings
and monopoles correspond to singular gauge potentials $A$. It is easy to imagine, therefore, that by
going to arbitrary large $A$, so to say inflated by the gauge transformations, one can generate 
an arbitrary number of spurious strings and monopoles. On the other hand, by minimizing potentials
one may hope to squeeze the number of the topological defects to its minimum and these topological
defects may have physical significance. In order to quantify what is understood by minimizing the
potentials, consider the Lorenz gauge which is defined by the requirement that the functional 
\beq
\label{R}
R~=~ \sum\limits_{x,\mu} (\, 1 - 1/2 \tr g_{x,\mu} \,)
\eeq
is minimal on the gauge orbit. In the naive continuum limit Eq.~(\ref{R}) reduces to $R=1/4\,\int (A^a_\mu)^2$.
Thus, the Lorenz gauge is singled out since in this gauge the link matrices are as close to unity as possible
and the gauge dependent singularities are suppressed as much as possible. As we shall see in a moment,
this heuristic justification of the gauge choice can be checked a posteriori. 

We have performed numerical experiments with monopoles defined by Eq.~(\ref{j}) in
the Lorenz gauge. The calculations were done on the $12^4$ and $4\mathrm{~x~}12^3$ lattices 
with periodic boundary conditions
using $30$ well equilibrated and statistically independent configurations.
We used the local over-relaxation algorithm to fix the Lorenz gauge. The gauge was considered
fixed when at each site the gauge transformation matrix $\Omega_x$,
which locally maximizes Eq.~(\ref{R}), satisfy  $1-1/2 \tr \Omega_x ~<~ 10^{-6}$. 
In order to circumvent the  Gribov copies problem, each thermalized configuration was randomly gauge transformed
to five gauge equivalent configurations and the Lorenz gauge was fixed again on each gauge copy.
The measurements were performed on the configuration for which the functional (\ref{R}) is minimal.

The density of monopoles, Eq.~(\ref{j}), is given by:
\beq
\label{density}
\rho ~=~ {1\over 4V}\sum\limits_{x,\mu} |j_{x,\mu}|\,,
\eeq
where $V$ is the lattice volume. We have measured the density $\rho$ in the Lorenz gauge as
a function of the bare coupling $\beta=4/g^2$. The logarithm of $\rho$ versus $\beta$ on the
symmetric $12^4$ lattice is shown on the Fig.~1a. The solid curve on the figure is the
renormalization group prediction:
\beq
\label{density-RG}
\rho ~=~ \mathrm{~const.~}\cdot\; \beta^{153/121}\;
\exp\left\{ - {9\pi^2 \over 11}\beta  \right\} \,,
\eeq
plotted in the logarithmic scale. As is clear from the plot, the density $\rho$ perfectly
scales towards the continuum limit. Thus the monopoles (\ref{j}) are physical objects and this
justifies the choice of the Lorenz gauge, motivated above.  Note that the finite physical
density of the monopoles (\ref{j}) in the continuum gluodynamics implies in particular, that
even in the Lorenz gauge the gauge fields are rather singular. The relevance of the singular
fields in QCD is discussed in Ref.~\cite{Consequences}.

It is interesting to consider also the behavior of the density (\ref{density}) across the
deconfinement phase transition. On the Fig.~1b we plot the logarithm of $\rho$ on the
$4\mathrm{~x~}12^3$ lattice, the solid curve is the scaling law (\ref{density-RG}).
The monopole density jumps at the critical coupling and remains non zero in the
deconfinement phase. Note that after the phase transition $\rho$ also seems to scale
correctly towards the continuum limit, although this question needs further investigations.

Another interesting quantity related to the monopole dynamics  is the asymmetry of the
monopole currents \cite{Asymmetry}:
\beq
\label{A}
A ~=~\frac{1}{3}\; <\sum\limits_{x,\mu=1,2,3} |j_{x,\mu}|> / <\sum\limits_{x} |j_{x,0}|>\,,
\eeq
which is known to be the order parameter of the deconfinement phase transition when
the monopoles are defined in the Maximal Abelian gauge. Indeed, the dual superconductor
confinement scenario suggests that the Abelian monopoles are condensed in the confinement phase,
while they are almost static at high temperatures. Therefore, the asymmetry of Abelian monopole
currents is unity at zero temperature and vanishes in the deconfinement phase with rising
temperature. Since the definition of the monopoles (\ref{j}) refers to the Lorenz gauge,
the behavior of the asymmetry (\ref{A}) across the phase transition is worth to be considered anew. 

The Fig.~2 represents the asymmetry $A$ versus $\beta$ on the $12^4$ (diamonds) and
$4\mathrm{~x~}12^3$ (squares) lattices. As expected, the asymmetry measured on the symmetric
lattice is unity within the numerical errors. On the asymmetric lattice the asymmetry shows
a rapid jump at the critical coupling, being substantially smaller after the phase transition.
Thus, in the deconfinement phase the monopoles defined in the Lorenz gauge, Eq.~(\ref{j}),
are mostly static. Note that the asymmetry needs not to be unity at $\beta < \beta_{crit}$ on
the asymmetric lattice since the Lorenz invariance is explicitly broken.

In order to clarify the physical relevance of the monopoles constructed consider the correlation
of the local SU(2) action density and the monopole currents \cite{BaChPo-97}. In particular,
we define the relative excess of the action density associated with the monopole as follows:
\beq
\label{eta}
\eta ~=~ { S_{m} ~-~ S  \over S }\,,
\eeq
where $S$ denotes the average vacuum action density $S=<1-1/2\tr U_p>$. The quantity $S_m$ is
the average value of $1-1/2\tr U_p$ calculated on the plaquettes, which belong to
the 3-dimensional cubes $C_{x,\mu}$ dual to the monopole currents $j_{x,\mu}$:
\beq
S_m ~=~ <{1\over 6} \sum\limits_{p\in \diff C_{x,\mu}} \left( 1 - {1\over 2}\tr U_p \right) >\,.
\eeq
The average is implied over all cubes $C_{x,\mu}$ for which $j_{x,\mu}\neq 0$. For static
monopoles only the magnetic part of the SU(2) action density contributes to $S_m$. We have
measured the quantity $\eta$ on the symmetric $12^4$ lattice, the result is plotted on the
Fig.~3.  It is clearly seen that the monopoles in the Lorenz gauge are indeed locally
correlated with non-Abelian SU(2) action density.

Such an excess of the action associated with monopoles is quite a common phenomenon. 
Indeed, monopoles are correlated with instantons \cite{instantons} which realistically
represent non-perturbative background in QCD \cite{instantons-review}. A possible general
connection between monopoles and background fields was recently pointed out in
Ref.~\cite{ChGuPoZa-00}.
Moreover, the excess of the action associated with monopoles (\ref{j}) is numerically close
to the action excess around the Abelian monopoles in the Maximal Abelian gauge \cite{BaChPo-97}.

The latter observation rises the question about the local correlation between these objects.
A priori, it is by no means evident that the monopoles defined by the Eq.~(\ref{j}) and the
Abelian monopoles in the Maximal Abelian gauge should be correlated.
Indeed, the monopoles in the Maximal Abelian gauge are by construction Abelian objects, 
the field of which have a natural electromagnetic interpretation. In particular, in order
to explain the confinement phenomenon as due to these Abelian monopoles, they should be
condensed in the vacuum of gluodynamics. Contrary to that, the Dirac strings and monopoles,
Eqs.~(\ref{lattice-Dirac-string},\ref{j}), are  non-Abelian field configurations. There is no
simple electromagnetic-like interpretation of them and therefore their relation to the confinement
is a separate question which should be considered anew. In particular, the contribution of the
field configurations which correspond the monopoles (\ref{j}) to the expectation value of the
Wilson loop is unknown. Moreover, one cannot say that if the monopoles (\ref{j}) are
related to the confinement, they should be condensed in the vacuum.
Therefore, the correlation of the Abelian monopoles in the Maximal Abelian gauge and the monopoles
defined by Eq.~(\ref{j}) is an interesting question, which we leave for future investigation.

Due to the non-Abelian nature of the strings (\ref{lattice-Dirac-string}) it is not evident a priori
that they are unphysical. Indeed, in case of Abelian Dirac strings the U(1) gauge freedom
guarantees that no physical result depends on the string position. This is true in particular
in Abelian gauges of gluodynamics. Contrary to that, the strings (\ref{lattice-Dirac-string})
are considered in the Lorenz gauge in which there is no remaining U(1) gauge symmetry.
Therefore the physical irrelevance of the strings (\ref{lattice-Dirac-string}) should
be checked separately.

We have numerically found that, indeed, the strings (\ref{lattice-Dirac-string}) are unphysical objects.
For example, their density, defined analogously to the Eq.~(\ref{density}),
scales at weak coupling according to the Eq.~(\ref{density-RG}), whereas the density of physically
relevant string-like excitations behaves as 
$\rho_{\mathrm{string}}\sim \exp\{ -\frac{6\pi^2}{11}\beta + \frac{102}{121}\ln\beta \}$.
Furthermore, the action density calculated around the string away from its boundary is the same
as the vacuum action density within the numerical errors. Thus the strings (\ref{lattice-Dirac-string})
are similar to the Dirac strings in the compact electrodynamics.
The work is currently in progress to find the rigorous explanation of this fact.

\subsection*{Conclusions.}
\noindent

We have considered the well known quantum mechanical notion of the geometrical Berry phase
in the context of non Abelian gauge theories. The bridge between the Berry phase and
non-Abelian gauge theories is provided by the Wilson loop which can be considered as 
a quantum mechanical evolution operator. The role of the corresponding time-dependent
Hamiltonian is played then by the gauge potential $A(t)$. This observation allows to
utilize in gauge theories results obtained earlier for the Berry phase in case of
Hamiltonians belonging to a Lie algebra of group $G$. In particular, it is known
that the space of physically distinct states in this case is the coset space $G/H$,
$H$ being the Cartan subgroup of $G$. Moreover, the evolution of this quantum mechanical
system is naturally described by the generalized coherent states.

We have applied these ideas to the Wilson loop operator in the SU(2) gluodynamics both in the
continuum and lattice formulations. In particular, we have presented an explicit construction
of the phase $\varphi$ of the Wilson loop, which is not restricted to the interval $[-\pi ; \pi]$,
but still satisfies $1/2 \tr W = \cos\varphi$. We have also shown that in the coherent states basis
the phase $\varphi$ naturally decomposes into dynamical and Berry phases and discussed the 
topological construction inherent to the Berry phase in gluodynamics. Our considerations, when
applied to an elementary plaquette, give a lattice analog of the Dirac strings and monopoles.
The U(1) group which is usually associated with monopole-like defects is now defined locally
in terms of the infinitesimal Wilson loops, or elementary plaquettes.

The division of the total phase into the dynamical and geometric phases is gauge dependent.
In case of quantum mechanics such a decomposition also depends on a particular choice of
the basis. We argue, therefore, that for physical applications one has to consider a
gauge in which the gauge fields are as smooth as possible, thus suppressing all gauge
dependent singularities. Following the logic similar to that of the paper in Ref.~\cite{GuStZa-99}
we identify this gauge as the Lorenz gauge.

We have studied the dynamics of the newly defined monopoles in numerical simulations on the lattice
and presented a strong evidences that these objects are relevant in the continuum limit.
Since in the continuum limit the field configurations which produce these monopoles become singular,
our findings support for the idea that even in the Lorenz gauge, in which the gauge potentials are as smooth
as possible, the singular fields are important. More generally, our results provide with a measure of
singular fields in the vacuum. Phenomenology related to such fields in the continuum gluodynamics
has been discussed in Ref.~\cite{Consequences} 

It is worth emphasizing that the construction we have presented is essentially non-Abelian. Only
in the particular case when all the gauge fields have the same color orientation,
it reduces to the known construction of Dirac strings and monopoles in the compact U(1) gauge model. 
Therefore, the relevance of our considerations to the confinement
requires further investigation. It is also important to study the relation, if any, between monopoles
we have constructed and other topological defects, for example, the Abelian monopoles in the
Maximal Abelian gauge of the gluodynamics.

\subsection*{Acknowledgments.}
\noindent
We acknowledge thankfully the fruitful discussions with M.N.~Chernodub, M.I.~Polikarpov and
L.~Stodolsky. The work of F.V.G. was partially supported by grants RFBR~99-01230a, RFBR~96-1596740
and INTAS~96-370.


\newpage
\thispagestyle{empty}
\begin{figure}[ht]
  \begin{minipage}{1.0\textwidth}
    \centerline{
      \psfig{file=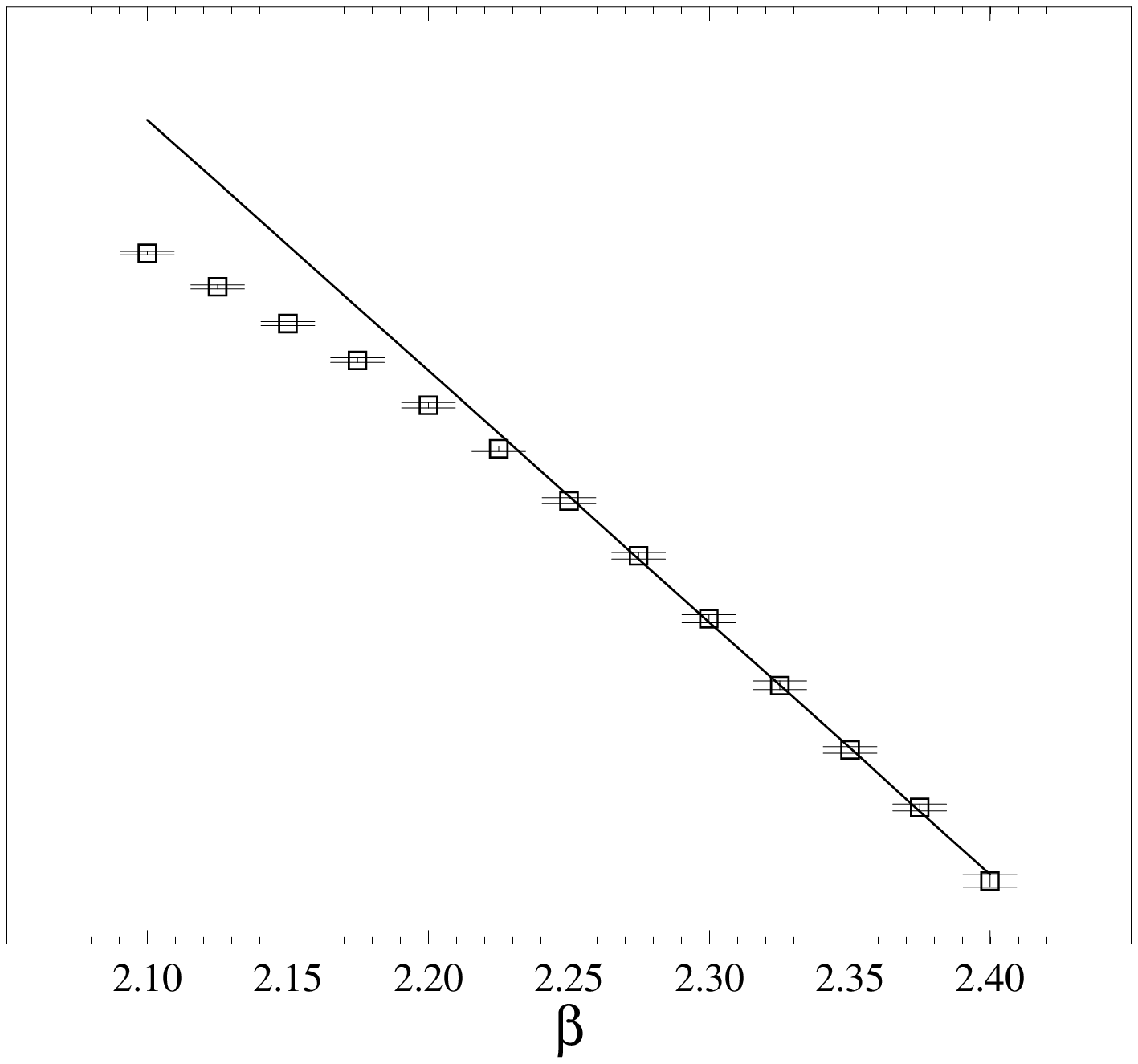,width=0.45\textwidth,silent=}
       \hspace{0.05\textwidth}
      \psfig{file=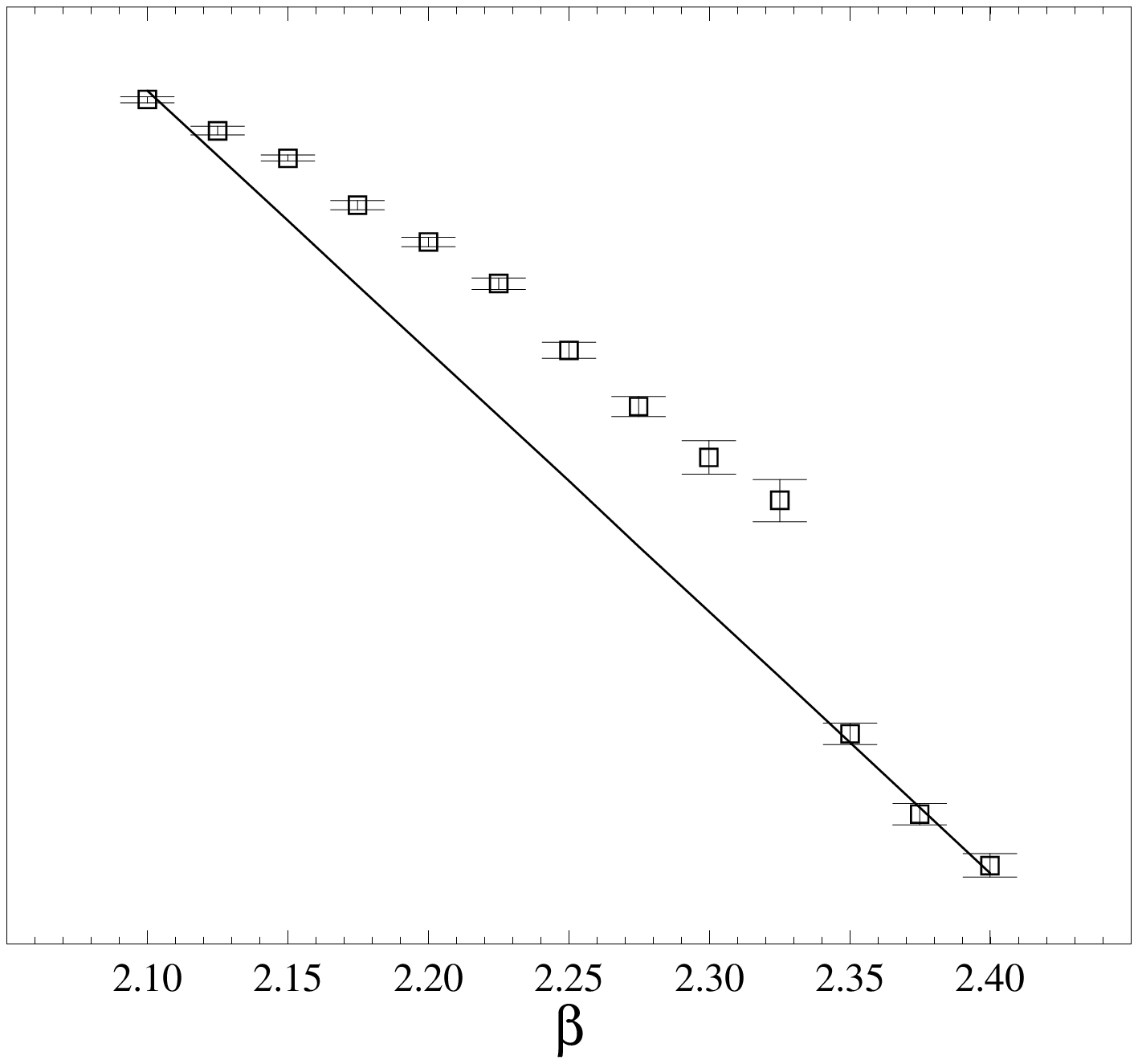,width=0.45\textwidth,silent=}
    }
    \centerline{(a) \hspace{0.45\textwidth} (b)}
  \end{minipage}
  \caption{
    The logarithm of the monopole density, Eq.~(\ref{density}), as a function of $\beta=4/g^2$.
    (a) -- the symmetric $12^4$  lattice; 
    (b) -- the asymmetric $4\mathrm{~x~}12^3$  lattice.
    Solid curve represents the renormalization group prediction, Eq.~(\ref{density-RG}).
  }

  ~

  ~

  \begin{minipage}{0.45\textwidth}
    \centerline{
      \psfig{file=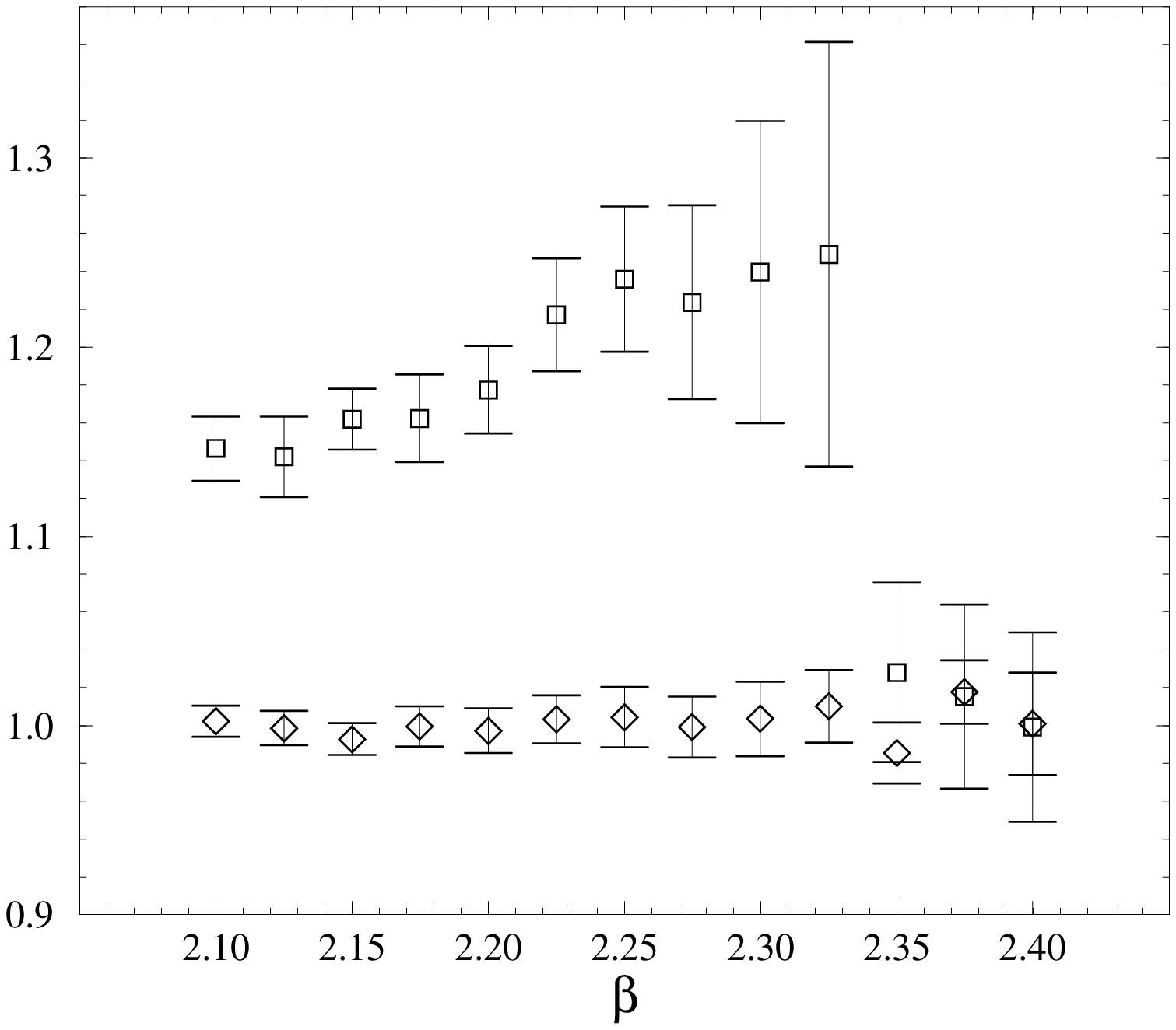,width=\textwidth,silent=}
    }
  \caption{
    The asymetry of the monopole currents, Eq.~(\ref{A}), versus $\beta$ on the $12^4$ (diamonds)
    and $4\mathrm{~x~}12^3$ (squares) lattices.
  }
  \end{minipage}
  \hspace{0.05\textwidth}   
  \begin{minipage}{0.45\textwidth}
    \centerline{
      \psfig{file=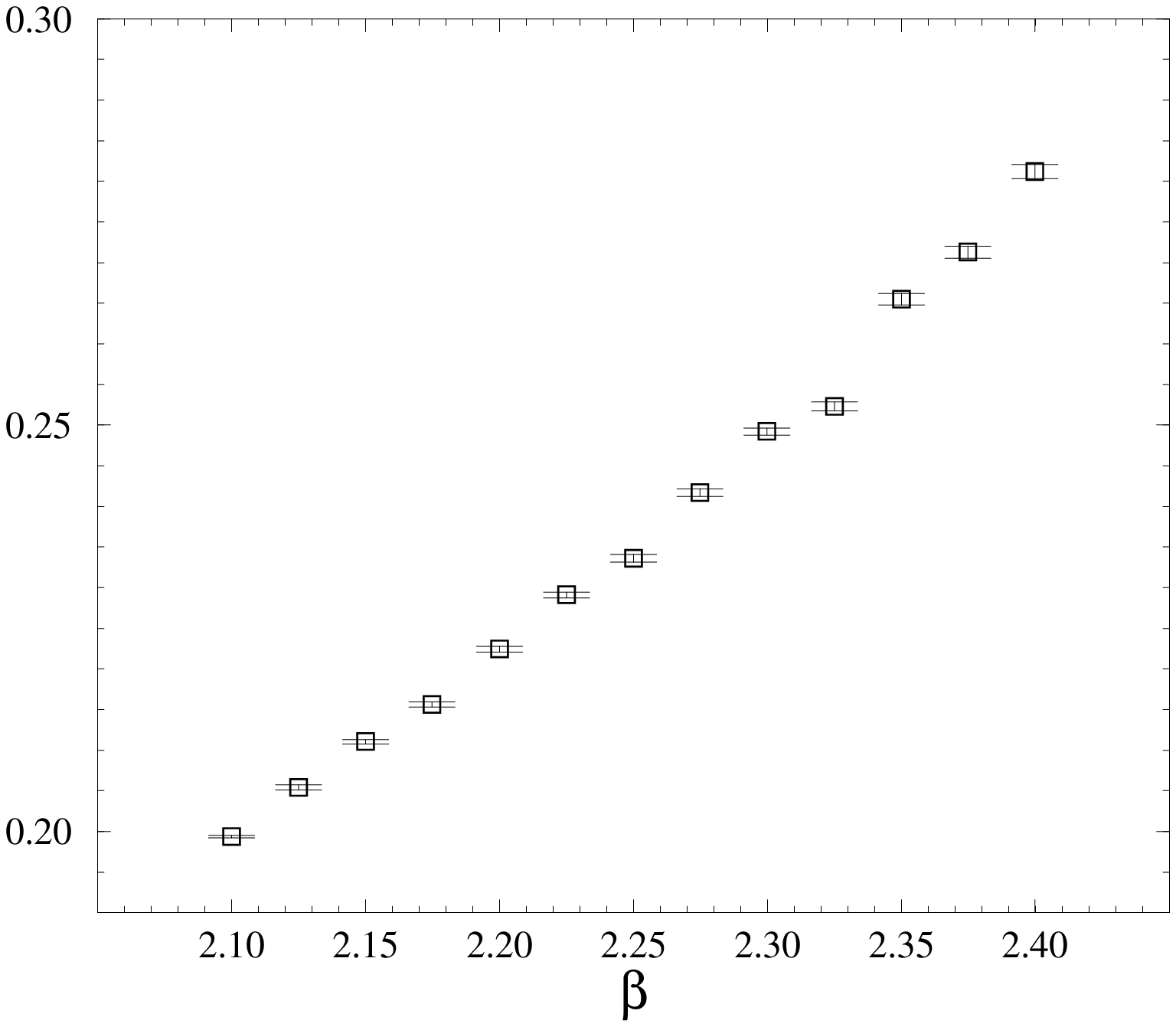,width=\textwidth,silent=}
    }
  \caption{
    The relative access of the magnetic action density, Eq.~(\ref{eta}),
    around the monopole currents, Eq.~(\ref{j}), for the $12^4$ lattice.
  }
  \end{minipage}
\end{figure}
\end{document}